\documentclass[epjCONF,columns]{svjour}
\usepackage{color}
\usepackage{graphics}
\usepackage{amsmath}
\usepackage[varg]{txfonts} 
\usepackage[latin1]{inputenc}

\session-title{Hadron Collier Physics (HCP) 2011, Paris, France}
\begin{document}
\def \et {E_{T}}
\def \pt {p_{T}}

\def  \met {\not\!\!\et }
\def \ETmiss {{E}_{T}^{miss}}
\newcommand{\fbinv}{\ensuremath{\textrm{fb}^{-1}}}
\newcommand{\lumifinal}{2.1\fbinv}
\newcommand{\TeV}{\ensuremath{\textrm{TeV}}}
\newcommand{\GeV}{\ensuremath{\textrm{GeV}}}
\newcommand{\eepm}{\Pep\Pem}
\newcommand{\mmpm}{\Pgmp\Pgmm}
\newcommand{\empm}{\ensuremath{\Pe^\pm \Pgm^\mp}}
\newcommand{\zjets}{\ensuremath{Z+\textrm{jets}}}
\newcommand{\ttbar}{\ensuremath{\textrm{t}\bar{\textrm{t}}} }
\newcommand{\mc}{Monte Carlo}
\newcommand{\JZB}{\ensuremath{\textrm{JZB}}}
\newcommand{\chiz}{\ensuremath{\tilde{\chi}^{0}}}
\newcommand{\sGlu}{\ensuremath{\tilde{g}}}
\newcommand{\LSP}{\ensuremath{\tilde{\chi}^{0}_{1}}}

\definecolor{SFZP}{rgb}   {0.0 , 0.0 , 0.0} 
\definecolor{OFZP}{rgb}   {0.6 , 0.0 , 0.0}
\definecolor{OFSB}{rgb}   {0.0 , 0.6 , 0.0}
\definecolor{SFSB}{rgb}   {0.0 , 0.0 , 0.6}
\definecolor{Bpred}{rgb}  {1.0 , 0.0 , 0.0} 

\newcommand{\Bpred}{\ensuremath{{\color{Bpred}JZB^\mathrm{pred}_\mathrm{bkgd}}}}
\newcommand{\SFZPJZBPOS}{\ensuremath{{\color{SFZP} \textrm{JZB}^\mathrm{SFZP}_\mathrm{pos}}}} 
\newcommand{\SFZPJZBNEG}{\ensuremath{{\color{SFZP} \textrm{JZB}^\mathrm{SFZP}_\mathrm{neg}}}} 
\newcommand{\OFZPJZBNEG}{\ensuremath{{\color{OFZP} \textrm{JZB}^\mathrm{OFZP}_\mathrm{neg}}}}
\newcommand{\OFZPJZBPOS}{\ensuremath{{\color{OFZP} \textrm{JZB}^\mathrm{OFZP}_\mathrm{pos}}}}
\newcommand{\SFSBJZBNEG}{\ensuremath{{\color{SFSB} \textrm{JZB}^\mathrm{SFSB}_\mathrm{neg}}}}
\newcommand{\SFSBJZBPOS}{\ensuremath{{\color{SFSB} \textrm{JZB}^\mathrm{SFSB}_\mathrm{pos}}}}
\newcommand{\OFSBJZBNEG}{\ensuremath{{\color{OFSB} \textrm{JZB}^\mathrm{OFSB}_\mathrm{neg}}}}
\newcommand{\OFSBJZBPOS}{\ensuremath{{\color{OFSB} \textrm{JZB}^\mathrm{OFSB}_\mathrm{pos}}}}

\newcommand{\SFZPJZB}{\ensuremath{{\color{SFZP} \textrm{JZB}^\mathrm{SFZP}}}} 
\newcommand{\OFZPJZB}{\ensuremath{{\color{OFZP} \textrm{JZB}^\mathrm{OFZP}}}}
\newcommand{\SFSBJZB}{\ensuremath{{\color{SFSB} \textrm{JZB}^\mathrm{SFSB}}}}
\newcommand{\OFSBJZB}{\ensuremath{{\color{OFSB} \textrm{JZB}^\mathrm{OFSB}}}}

\title{Search for supersymmetry in events with a Z boson, jets and missing energy}
\author{Marco - Andrea Buchmann\inst{1}\fnmsep\thanks{\email{marco.andrea.buchmann@cern.ch}} on behalf of the CMS Collaboration}
\institute{Institute for Particle Physics, ETH Z\"urich, CH--8093 Z\"urich, Switzerland}
\abstract{
We present a search for Physics beyond the Standard Model (SM) in final states with a Z boson, jets and missing transverse energy, using a data sample collected in 2011 by the CMS detector at the Large Hadron Collider 
corresponding to an integrated luminosity of 2.1 fb$^{-1}$. This final state is predicted in several models of Physics beyond the SM, including supersymmetry. A novel analysis method is exploited, the Jet-Z Balance 
method, and a precise determination of the total SM background is obtained using a control sample from data. In the absence of any significant excess beyond the SM background, upper limits are set on simple models 
of supersymmetry, and further information is provided to allow confrontation of other models to these results.
} 
\maketitle
\noindent We describe a search~\cite{JZB} for Physics beyond the Standard Model (BSM) in a sample
of pp collisions collected by the Compact Muon Solenoid (CMS) detector~\cite{CMS} at the Large Hadron Collider (LHC),
at a center-of-mass energy of $7~\textrm{TeV}$. The size of the data sample corresponds to~$2.1$~fb$^{-1}$.
We search for final states with a Z boson, jets and missing energy ($\ETmiss$) where the
Z boson decays to electron or muon pairs. 
This final state is a clean and distinct signature present in many models of BSM Physics,
in particular supersymmetry (SUSY)~\cite{Matchev:1999ft,Ruderman:2011vv}.

The most significant Standard Model (SM) backgrounds in this final state are \zjets\ processes and \ttbar\ production. The
former is the most challenging since it peaks in the mass distribution of the two leptons and can have apparent missing energy 
due to detector resolution and reconstruction effects.

The Jet-Z balance (JZB) method has been devised to predict the contribution from \zjets\ events~\cite{phdthesis}.
We use the flavor symmetry of \ttbar\ decays, as well as dilepton invariant mass side-band regions to
predict the contribution from \ttbar\ events. 

We select events with at least three central jets
with transverse momentum $\pt>30~\GeV$ and two opposite-sign leptons with $\pt>20~\GeV$.

\begin{figure}[t!]
  \begin{center}
\resizebox{0.90\columnwidth}{!}{
    \includegraphics{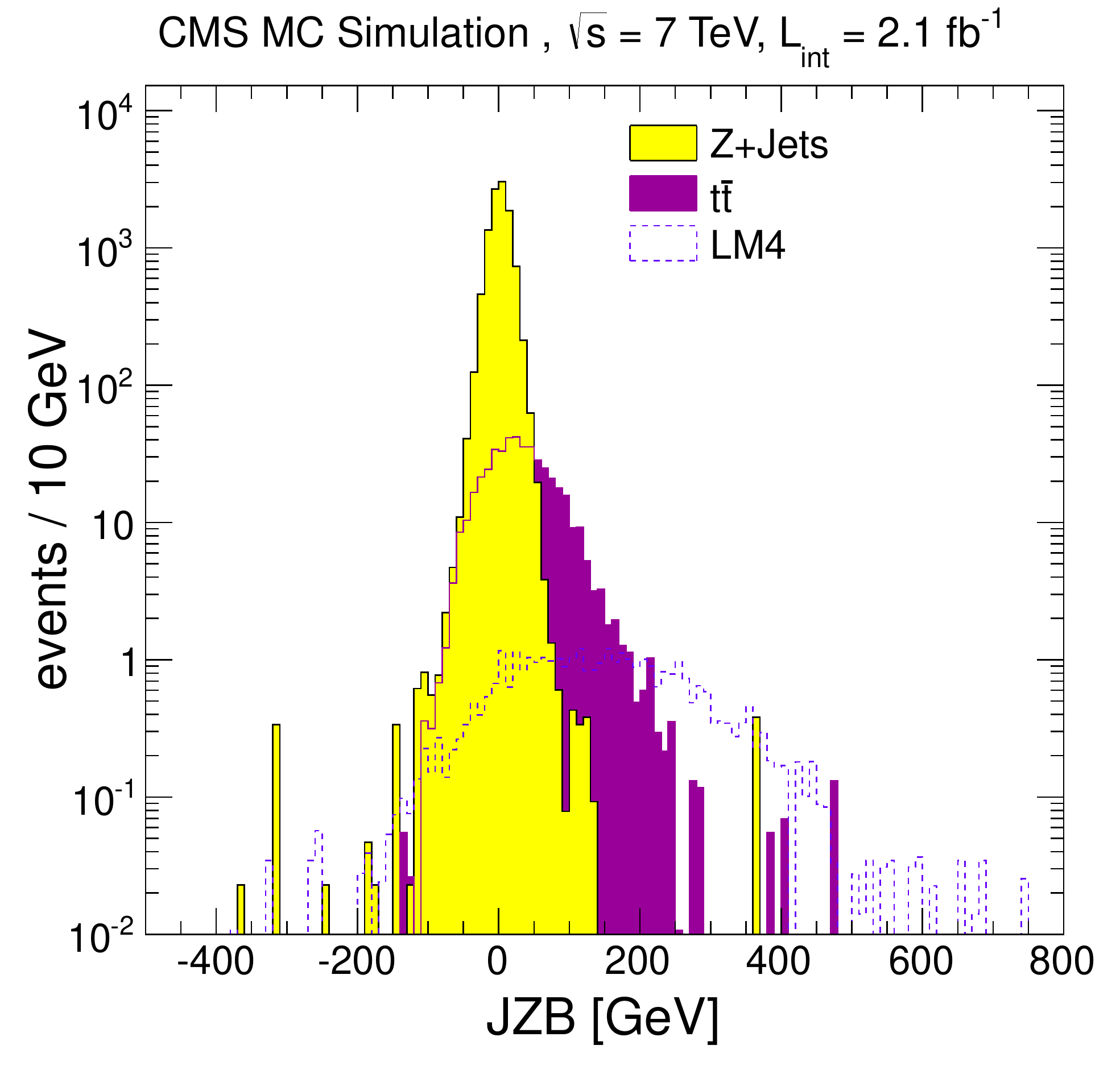}
}

    \caption{$\JZB$ distribution in $\zjets$, $\ttbar$ and signal MC simulation scaled to~$\lumifinal$,
     after preselection and dilepton invariant mass requirement.
    The signal distribution corresponds to the SUSY LM4 scenario.
    }
    \label{fig:mcjzb}  
  \end{center}
\end{figure}

The $\JZB$ variable is defined as the difference between the $\pt$ of the vectorial sum of all 
jets and the $\pt$ of the Z boson candidate: 

\begin{equation*}
JZB = |\sum_\textrm{jets}\vec{p}_{T}| - \left|\vec{p}_T^{(Z)}\right|
\end{equation*}
The JZB variable has a strong discriminating power between instrumental MET events (which populate the left and the right side evenly) and real MET events, like signal events, which are shifted to the right.
The JZB distribution is shown in Fig.~\ref{fig:mcjzb}.

A low-, a mid-, and a high-\JZB\ search region are defined by requiring $\JZB>50\GeV$,
$\JZB>100\GeV$, and $\JZB>150\GeV$, respectively. These requirements efficiently suppress 
the \zjets\ background, leaving \ttbar\ as the dominant background. 

The signal region, denoted \SFZPJZBPOS, consists of same-flavor events in which the invariant mass
of the two leptons is compatible with the Z mass hypothesis ($|m_{\ell\ell}-m_{Z}|<20~\GeV$)
and $\JZB>0~\GeV$. 
Three control regions are used to predict the contribution of flavor-symmetric backgrounds 
(mainly coming from \ttbar\ processes): 
\begin{itemize}
 \item opposite-flavor events compatible with the Z boson mass hypothesis (\OFZPJZB)
 \item opposite-flavor events in the side-band of the Z boson mass (\OFSBJZB) and
 \item same-flavor events in the same side-band region (\SFSBJZB)
\end{itemize}
The side-band region is defined as $55~\GeV<m_{\ell\ell}<70~\GeV \cup 112~\GeV<m_{\ell\ell}<160~\GeV$. 

\begin{figure*}[h!t]
  \begin{center}
\resizebox{0.83\columnwidth}{!}{
    \includegraphics{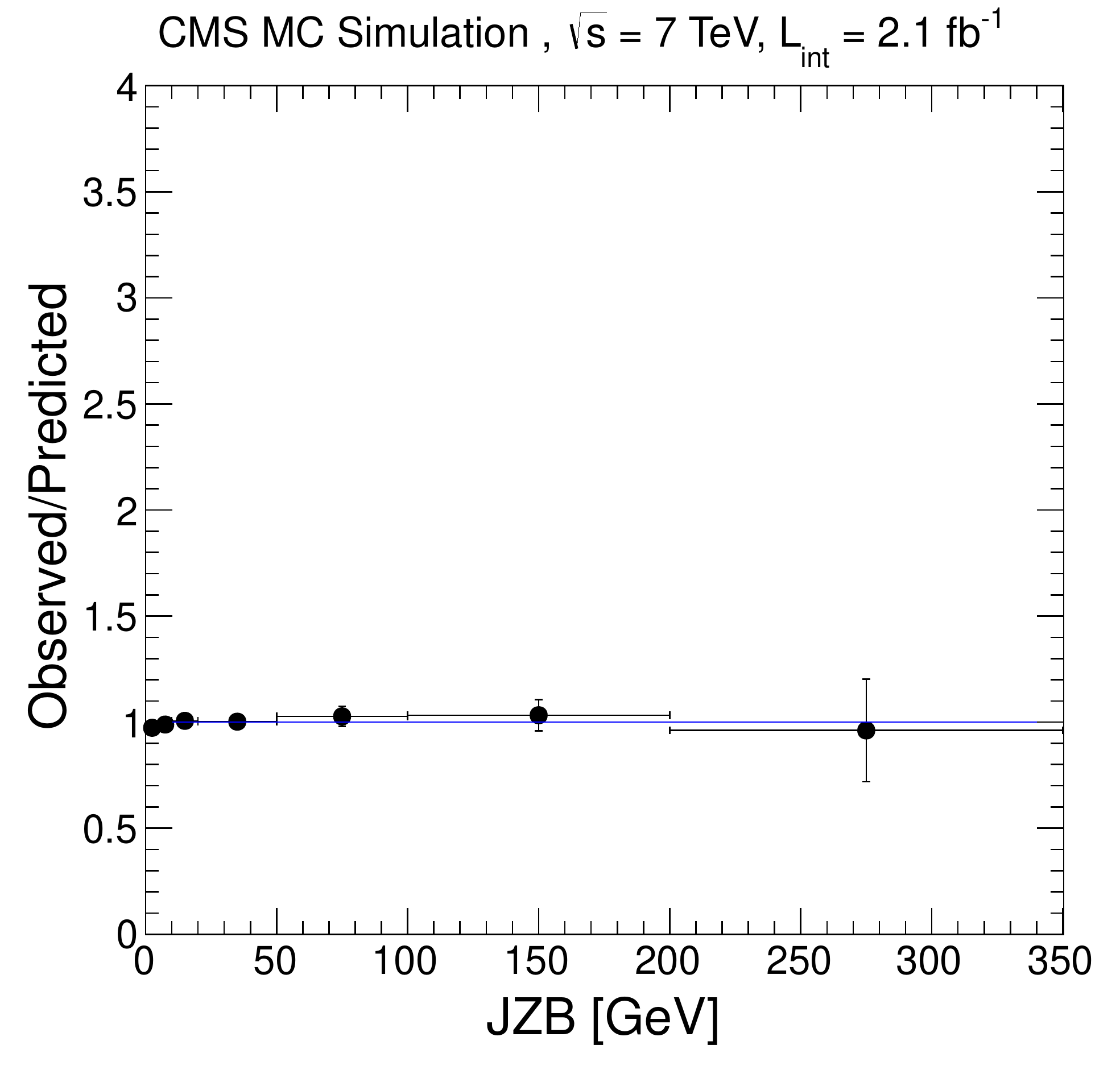}
}
\resizebox{0.83\columnwidth}{!}{
    \includegraphics{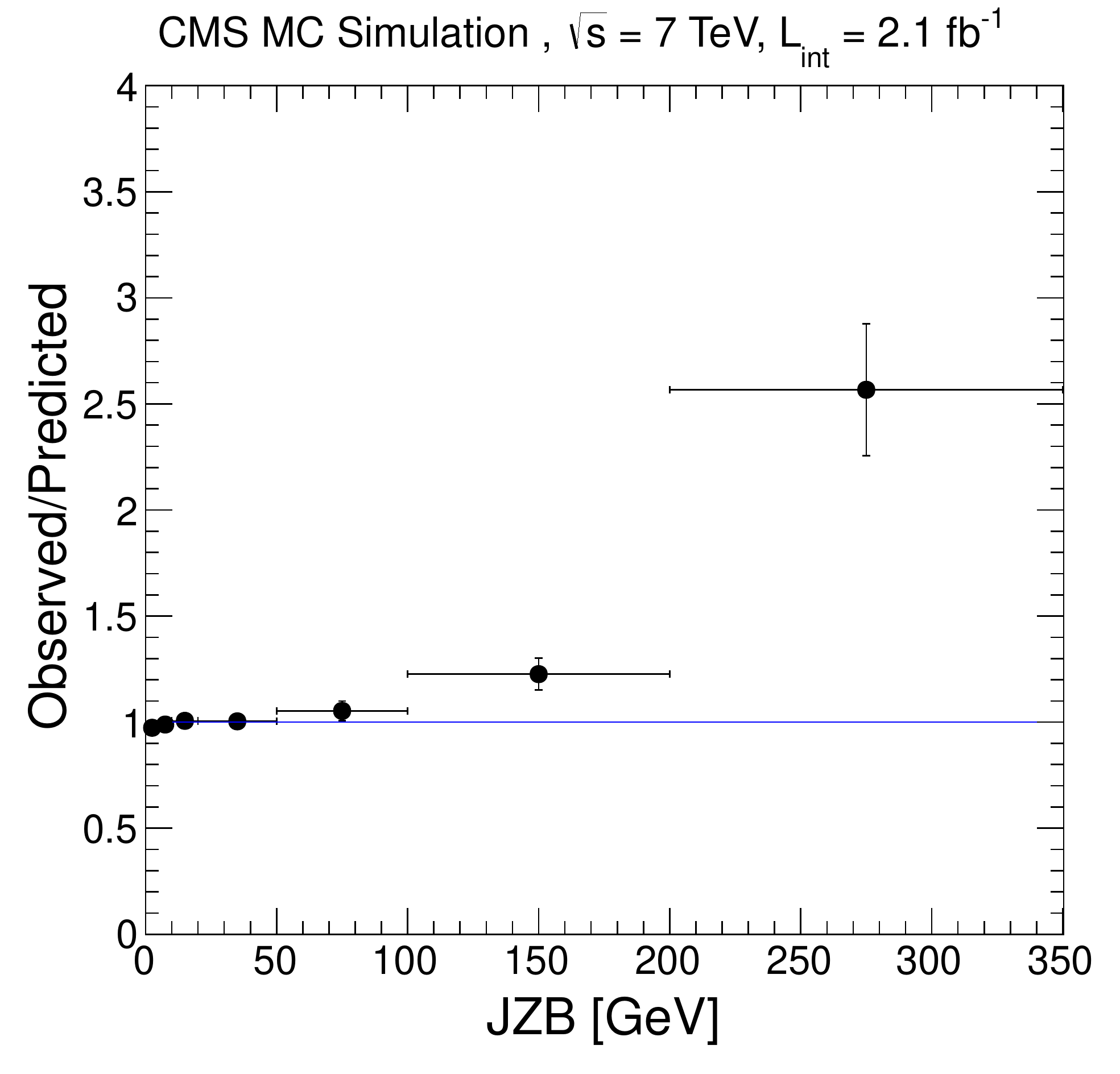}
}
    \caption{Ratio between observed and predicted \JZB\ distributions in \mc\ simulation,
    without (left) and with (right) the inclusion of signal \mc\ simulation.
    The errors correspond to the statistical uncertainty of the \mc\ simulation.}
    \label{fig:mcclosure}
  \end{center}
\end{figure*}

\begin{figure*}[h!bt]
  \begin{center}
\resizebox{0.83\columnwidth}{!}{
    \includegraphics{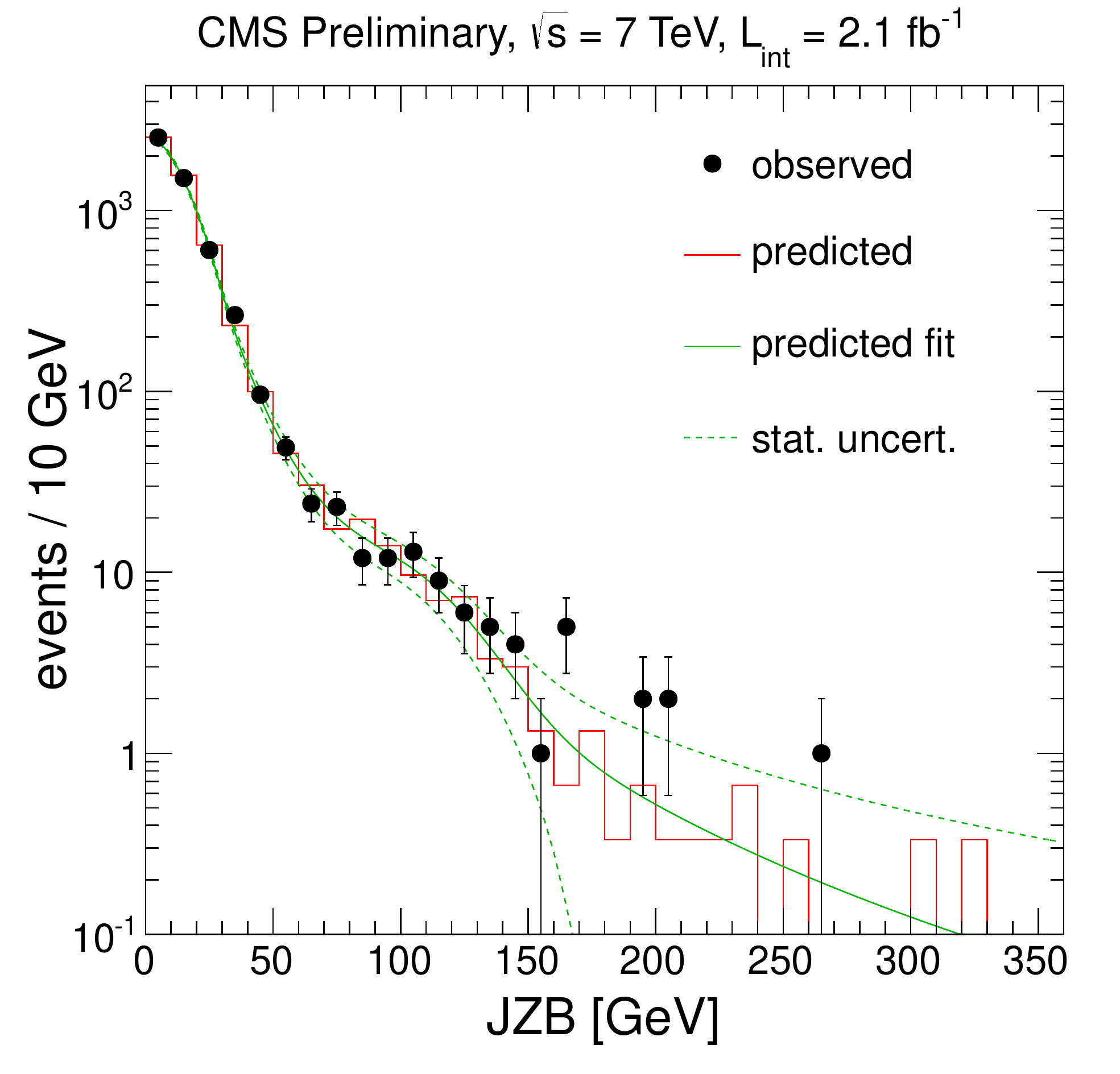}
}
\resizebox{0.83\columnwidth}{!}{
    \includegraphics{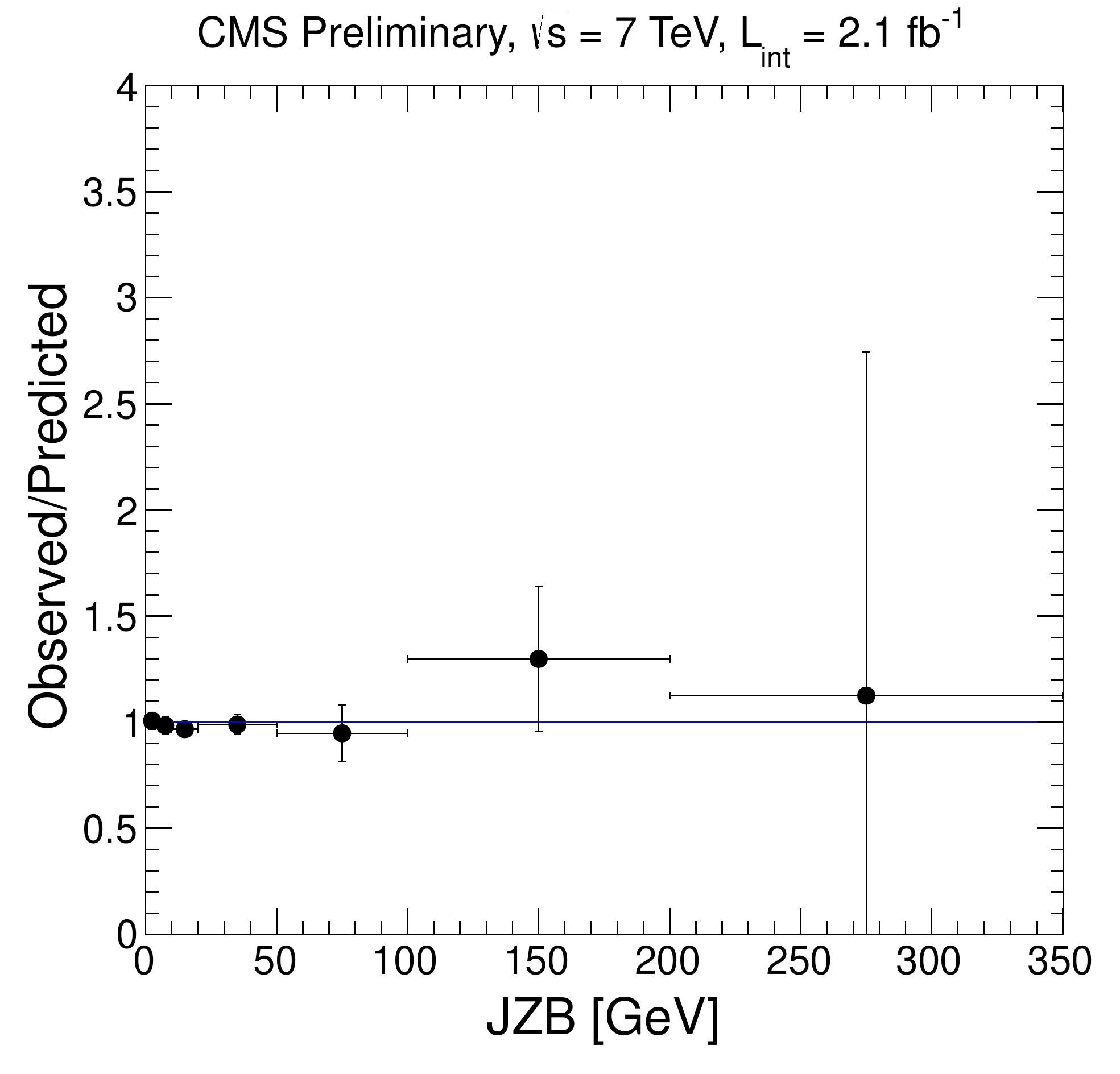}
}
    \caption{Comparison between the predicted and observed \JZB\ distributions in the search region.
    The predicted distribution has been fitted to display 1-$\sigma$ error bands.
    The right plot shows the ratio between the two distributions on the left.}
    \label{fig:Bpred}
  \end{center}
\end{figure*}

\newpage

The contribution from flavor-symmetric backgrounds is computed as the average of the three control regions,
as they provide an independent estimate of the same background process. The contribution from
the \zjets\ background is estimated from the \SFZPJZBNEG region after subtracting the contribution from flavor-symmetric
backgrounds in this region. The total background prediction in the signal region is then computed as:

\begin{align*}
\Bpred = |\SFZPJZBNEG| &+ \frac{1}{3}|\OFZPJZBPOS| - \frac{1}{3}|\OFZPJZBNEG|\\ 
       &+ \frac{1}{3}|\SFSBJZBPOS| -  \frac{1}{3}|\SFSBJZBNEG| \\
       &+ \frac{1}{3}|\OFSBJZBPOS| - \frac{1}{3}|\OFSBJZBNEG|
\end{align*}

We assign a systematic uncertainty of 25\% to \SFZPJZBNEG\ and \SFZPJZBPOS\ and a systematic uncertainty of 50\% to each of the estimates from the three control regions for flavor-symmetric backgrounds.
The background estimation methods are validated in \mc\ simulation, in a mixture of
all SM backgrounds, with and without the inclusion of signal \mc\ simulation. 
The ratio between observed and predicted distributions is shown in Fig.~\ref{fig:mcclosure} 
for the two cases. There is very good agreement in the background-only hypothesis and 
good sensitivity to a possible signal.

The comparison between predicted and observed distributions in data is shown in Fig.~\ref{fig:Bpred}. 
The observed and predicted yields in these regions are summarised in Tab.~\ref{tab:results}.

\begin{table}[hbtp]
\renewcommand{\arraystretch}{1.3}
  \begin{center}
  \caption{Total number of events observed in the search regions, 
  and corresponding background prediction.}\label{tab:results}
  \begin{tabular}{lcc}
  \hline
  Region        & Observed   & Background prediction   \\
  \hline
  $\JZB>50\GeV$  & 168 & $164  \pm10  \textrm{(stat)}\pm42 $(sys)    \\
  $\JZB>100\GeV$ &  48 & $ 37  \pm 4  \textrm{(stat)}\pm10 $(sys)   \\
  $\JZB>150\GeV$ &  11 & $  7.0\pm 1.5\textrm{(stat)}\pm 2.1 $(sys)   \\
  \hline
  \end{tabular}
  \end{center}
\end{table}

\begin{figure}[hbtp]
  \begin{center}
\resizebox{0.75\columnwidth}{!}{
    \includegraphics{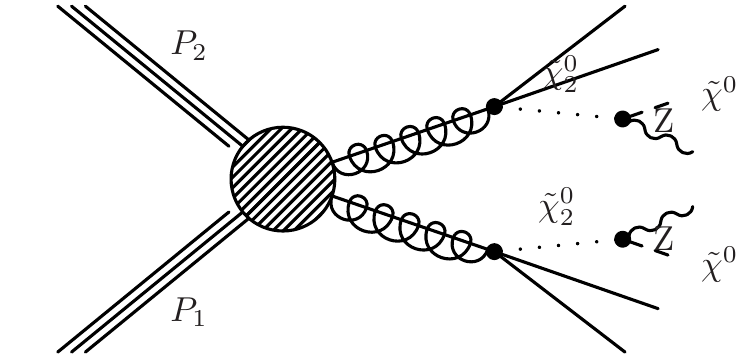}
}
    \caption{Simplified model for the production of two gluinos decaying into jets and a Z boson}
    \label{fig:t5zz}
  \end{center}
\end{figure}

In the absence of any significant excess, we set upper limits
on the production cross-section of simplified models of SUSY (SMS). 
In the process considered in this study (shown in Fig.~\ref{fig:t5zz}), two gluinos are generated, 
decaying to jets and a neutralino, which in turn decays to a Z boson and the lightest SUSY particle (LSP). 
The parameters of the model are the masses of the gluino ($m_{\tilde{g}}$) and the LSP ($m_{\chiz_{1}}$).
The mass of the intermediate neutralino ($m_{\chiz_{2}}$) is fixed to the mean of the two other masses.
Signal contamination is fully accounted for and the obtained limits are reliable.

\begin{figure}[hbtp]
  \begin{center}
\resizebox{0.83\columnwidth}{!}{
    \includegraphics{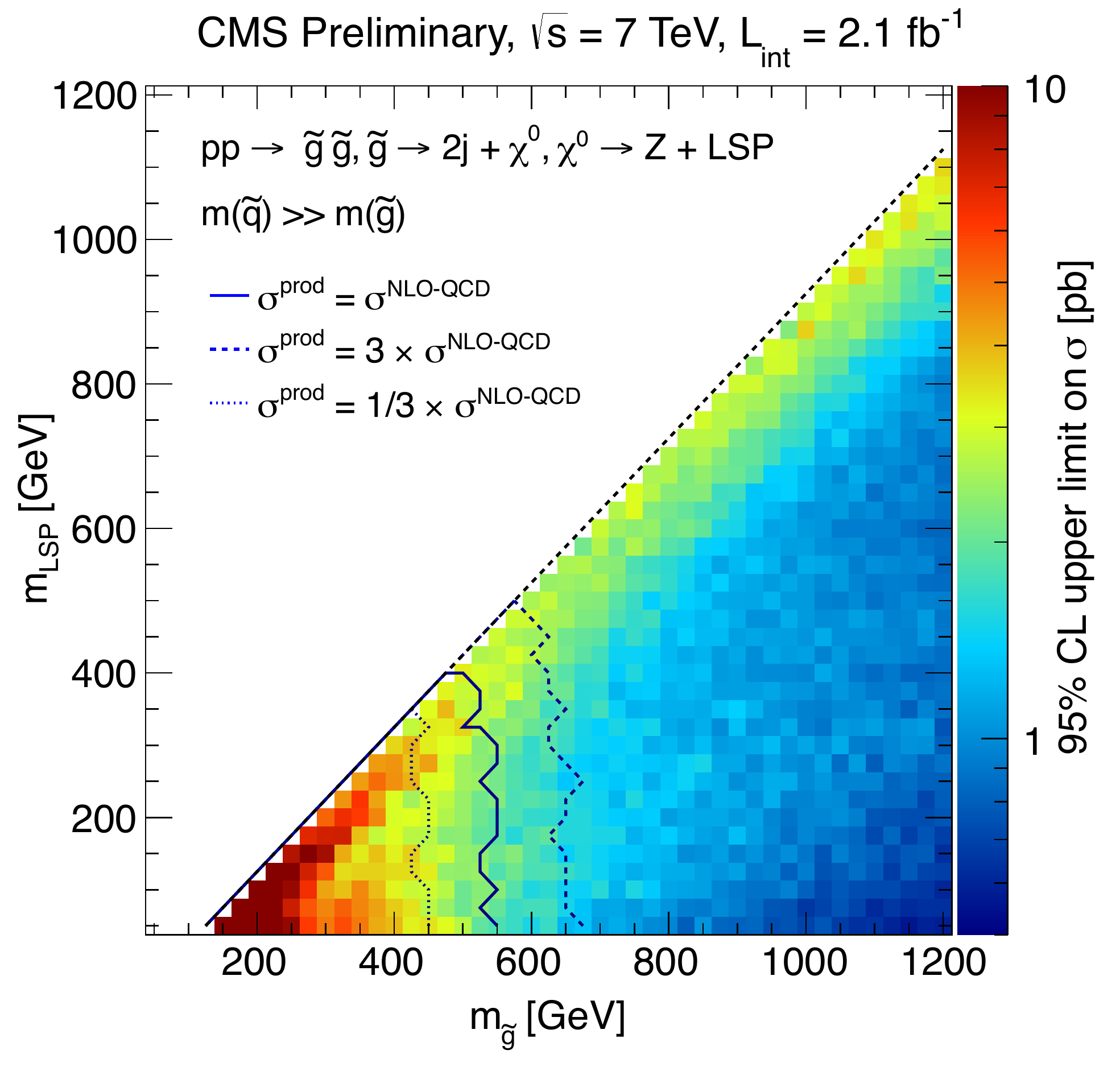}
}
    \caption{95\% CL upper limits on the cross-section of the inclusive Z boson decay mode
    in the $(m_{\sGlu},m_\text{LSP})$ SMS parameter space. In each point, the best expected limit from the 
    low-, mid-, and high-\JZB\ regions is used. The exclusion curves for reference cross-sections are also shown in this plot.
}
    \label{fig:SMSlimits}
  \end{center}
\end{figure}

The 95\% CL upper limits on the cross-section for the topology described above are shown in Fig.~\ref{fig:SMSlimits}.
The upper limits are computed using a frequentist CL${}_\text{S}$ method~\cite{CLS}.
In order to interpret these limits in terms of gluino pair production cross-section, we use a reference cross-section $\sigma_{\text{ref}}$
and draw the 95\% CL exclusion contours on $\tfrac{1}{3}$, 1 and 3 times $\sigma_{\text{ref}}$. The reference cross-section 
corresponds to gluino pair production in the limit of infinitely heavy squarks, calculated at next to leading order
using PROSPINO~\cite{Beenakker-1996} and CTEQ6~\cite{Pumplin:2002vw} parton distribution functions.

In this simplified model, the \JZB\ analysis is mostly sensitive to topologies in which the Z boson and the LSP 
have a momentum in the rest frame of the neutralino that is small with respect to the momentum of the neutralino~\cite{JZB2010}. 
In the considered set of simplified models, this topology is realized in the region where the difference between the LSP mass and 
the gluino mass is small, as the JZB distribution for signal is very asymmetric and signal contamination has little impact on the sensitivity.

We also estimate upper limits in the context of two CMSSM benchmark scenarios, the LM4 and LM8 scenarios (see Tab.~\ref{tab:lmresults}).
LM4 (and LM8) are defined as $m_{0} = 210\ (500)\GeV$, $m_{1/2}=285\ (300)\GeV$, $\tan\beta=10$, sign($\mu) = +$, 
and $A_{0}=0\ (-300)\GeV$, respectively. 
The LM4 scenario is excluded at 95\% CL by our search.

\begin{table}[hbtp]
\renewcommand{\arraystretch}{1.3}
\setlength{\belowcaptionskip}{6pt}
\begin{center}
\caption{Observed upper limits on the cross section of LM4 and LM8 benchmark points for different cuts on $\JZB$.
The last column indicates the NLO cross-section of the two scenarios.}\label{tab:lmresults}
\begin{tabular}{  l   c  c c c }
 \hline
Scenario & $>50\GeV$  & $>100\GeV$ & $>150\GeV$ & $\sigma$ \\ 
\hline
LM4      & 7.4 pb         & 3.8 pb         & 1.9 pb        & 2.53 pb      \\ 
LM8      & 7.9 pb         & 4.2 pb         & 2.0 pb        & 1.03 pb      \\ 
\hline
\end{tabular}
\end{center}
\end{table}
We have presented a search for physics beyond the SM in $Z+\textrm{jets}+\ETmiss$ events using a data sample corresponding to an integrated luminosity of \lumifinal.
We do not observe any deviation between the observed number of events in the signal region and
the predicted number of events from data control samples. In the absence of any hint of Physics beyond the SM,
we interpreted our result using simplified models of SUSY.

\end{document}